# A Survey on Security and Privacy Issues in Edge Computing-Assisted Internet of Things

Abdulmalik Alwarafy, Khaled A. Al-Thelaya, Mohamed Abdallah, *Senior Member, IEEE,* Jens Schneider, and Mounir Hamdi, *Fellow Member, IEEE*

*Abstract*—Internet of Things (IoT) is an innovative paradigm envisioned to provide massive applications that are now part of our daily lives. Millions of smart devices are deployed within complex networks to provide vibrant functionalities including communications, monitoring, and controlling of critical infrastructures. However, this massive growth of IoT devices and the corresponding huge data traffic generated at the edge of the network created additional burdens on the state-of-the-art centralized cloud computing paradigm due to the bandwidth and resources scarcity. Hence, edge computing (EC) is emerging as an innovative strategy that brings data processing and storage near to the end users, leading to what is called EC-assisted IoT. Although this paradigm provides unique features and enhanced quality of service (QoS), it also introduces huge risks in data security and privacy aspects. This paper conducts a comprehensive survey on security and privacy issues in the context of EC-assisted IoT. In particular, we first present an overview of EC-assisted IoT including definitions, applications, architecture, advantages, and challenges. Second, we define security and privacy in the context of EC-assisted IoT. Then, we extensively discuss the major classifications of attacks in EC-assisted IoT and provide possible solutions and countermeasures along with the related research efforts. After that, we further classify some security and privacy issues as discussed in the literature based on security services and based on security objectives and functions. Finally, several open challenges and future research directions for secure EC-assisted IoT paradigm are also extensively provided.

## I. INTRODUCTION

INTERNET of Things (IoT) refers to a collection of things such as smart devices, sensors, actuators, or anything embedded with electronics that are connected through the Internet to send, store and receive data relevant to a particular service or application [1], [2]. The explosive progress of information technology enables IoT to support and boost the arrival of new innovative services and applications. Furthermore, IoT smart devices are continuously equipped with advanced and sophisticated sensing, computation, and processing power capabilities, which make them deployable in various complex environments. Fig. (1) shows some common IoT services and applications deployed in various vital sectors. According to a report from the International Data Corporation (IDC) [3], [4],

The authors are with the Division of Information and Computing Technology, College of Science and Engineering, Hamad Bin Khalifa University, Qatar (e-mail: aalwarafy@hbku.edu.qa; kalthelaya@hbku.edu.qa; moabdallah@hbku.edu.qa; jeschneider@hbku.edu.qa; mhamdi@hbku.edu.qa).

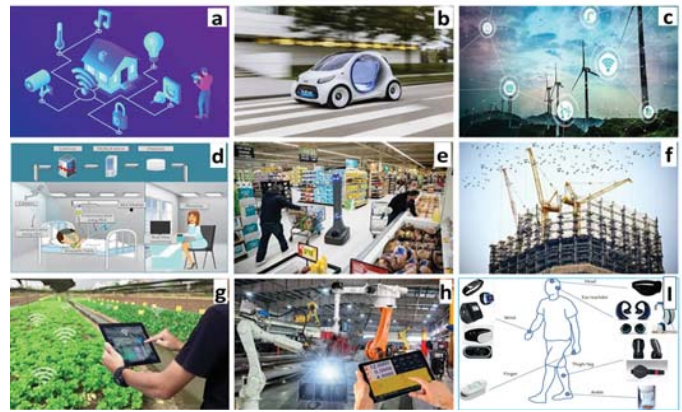

Figure 1: Applications of IoT. (a) smart buildings, (b) smart vehicles, (c) energy management, (d) health monitoring, (e) food supply chain, (f) construction management, (g) environmental monitoring, (h) production management, and (I) wearable devices.

the total number of connectable IoT smart devices/sensors, such as smartphones/tablets, smart home appliances, wearable devices, etc., is expected to exceed 200 billion by 2020, 30 billions of them will be indeed connected to the Internet. Such devices/sensors will produce and collect a tremendous amount of data from the surrounding environment, which is expected to exceed 500 Zettabytes (ZB) by 2020, according to a report from Cisco Global Cloud Index (GCI) [5]. In the standard cloud computing paradigm, all this data will be migrated to the sophisticated central servers located at the cloud for further processing, computation, and/or storage. The post-processed data needs then to be sent back to the end devices. Such a mechanism creates extra burdens on the core network as well as provides a poor quality of service (QoS), due to the following reasons: 1) there are extra costs in the data transmission due to the under-utilization of bandwidth and resources, 2) the increase in data size will drastically decrease network performance, 3) the explosive growth in the number of IoT devices will make it quite difficult to manage network connectivity and traffic, and 4) time-sensitive IoT services and applications including smart transport, smart electricity grid, and smart city, will suffer from unacceptable long delays. All these issues and limitations can be efficiently alleviated by adopting the edge computing-assisted IoT architecture. In such architecture, we combine the current cloud computing infrastructure with the edge computing (EC) paradigm to efficiently address the aforementioned problems.

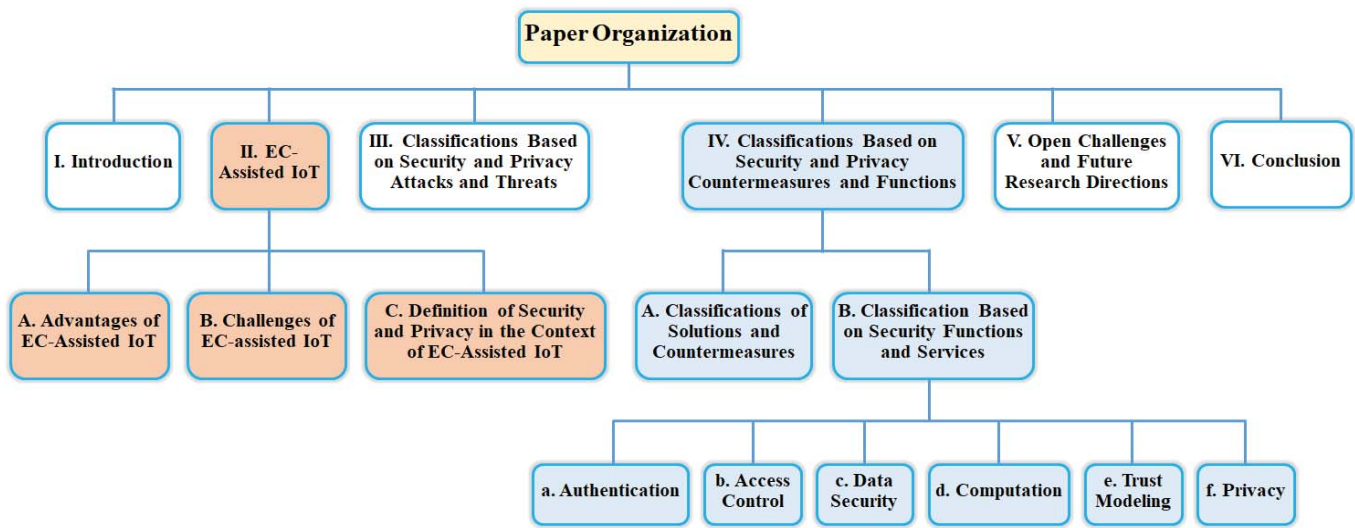

Figure 2: Paper organization.

This is achieved by locating nodes/servers near the network edge, closer to data sources [6]–[9]. Doing so will support IoT services and applications with reduced latency, flexible access, and enhanced network security. According to the IDC [3], the network edge will be responsible for processing and storing 40% of edge-originated data in the future EC-assisted IoT architecture.

As illustrated in Fig. (1), EC-assisted IoT systems are involved in managing and controlling a massive amount of data related to vital and sensitive applications in different sectors ranging from health monitoring to smart buildings. This has made it a target for attacks including hacking, cybercriminals, and governmental attacks. Adversaries may hack IoT devices/sensors to steal sensitive information such as financial accounts, bank cards, location data, and health information. Attackers may also spy on individuals or even launch protest campaigns against an organization. Furthermore, it is reported in [2] that more than 25% of the botnet attacks were originated from IoT devices, including home appliances, baby monitors, and smart TVs. Moreover, many websites in 2016, such as Netflix, Twitter, and Spotify, have been attacked by an organized distributed denial of service attacks originated from IoT smart devices. Therefore, it is crucial to conduct extensive and in-depth studies and develop effective solutions to handle security and privacy threats in the EC-assisted IoT networks. This would enable the development of secure smart devices/sensors for the emerging EC-assisted IoT services and applications.

There are several published research works aimed at addressing the aforementioned issues. Some of these papers are surveys related to the security of IoT in general without considering the EC aspect [1], [2], [10]–[12], while other papers are proposing and developing security and privacy-related solutions and countermeasures for EC-assisted IoT [8], [9], [13]–[34]. Although there are existing surveys related to security and privacy in the context of EC-assisted IoT [5]–[7], [19], [35]–[42], they are either; 1) still missing some of the most recent and prominent research works, 2) covering a limited number of security and privacy issues, 3) do not adequately cover the security and privacy attacks along with their countermeasure, 4) just presenting particular case studies for specific operating scenarios, or 5) considering different aspects of classifications. Motivated by the aforementioned security and privacy issues, and the research gaps and scarcity of existing literature in the context of EC-assisted IoT, this paper is proposed to fill these gaps and to overcome these shortcomings. In particular, this paper provides a comprehensive literature survey on security and privacy issues in the context of EC-assisted IoT. The main contributions of this paper are summarized as follows:

- We provide an overview of the EC-assisted IoT paradigm, including definitions, applications, and architecture. We also describe the advantages and limitations of EC-assisted IoT systems. Then, we define security and privacy in the context of EC-assisted IoT.
- We present thorough classifications of attacks and threats. Then, we discuss the possible solutions and countermeasures at different network layers and for different security and privacy issues. We also summarize some of the most recent research efforts pertaining to security and privacy in the context of EC-assisted IoT. Hence, the reader will be provided with an in-depth analysis of which attacks have been launched, what countermeasures have been considered in the literature to address them, and which threats still lurk.
- We extract, analyze, and summarize the most prominent security and privacy issues of EC-assisted IoT as reported in the literature. We also classify them based on EC-assisted IoT services and based on security objectives and functions.
- We extensively outline and describe some security and privacy-related open challenges, and provide deep insights into some promising future research directions in the context of EC-assisted IoT paradigm.

The rest of this paper is organized as follows. Section II provides a background related to the EC-assisted IoT paradigm. Definitions, applications, and architecture of this technology are described. Section III gives classifications of security and privacy attacks and threats for EC-assisted IoT. Section IV describes the possible security solutions and countermeasures. It also gives a comprehensive analysis on security and privacy issues for EC-assisted IoT. Classifications based on EC-assisted IoT services and based on security objectives and functions are also provided in Section IV. Section V provides open challenges along with future research directions. Finally, section VI summarizes the paper. The organization of the paper is illustrated in Fig. (2).

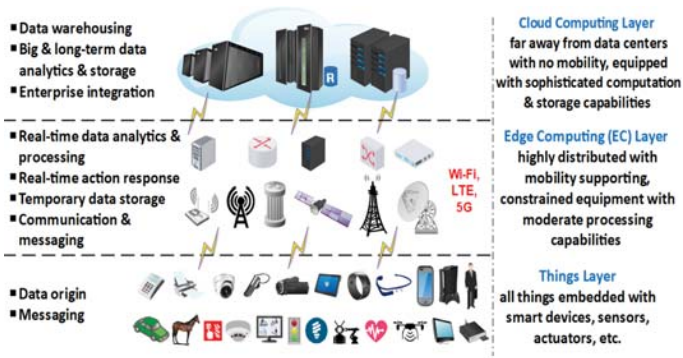

Figure 3: Standard layer architecture of EC-assisted IoT.

## II. Integration of Edge Computing and IoT: EC-Assisted IoT

This section provides an overview of the fundamental concepts, applications, and architecture of the integrated IoT and EC paradigm. Related research efforts will be also cited.

Both IoT and EC are separately rapidly evolving. Nevertheless, the characteristics of each paradigm are quite similar [35]. Therefore, IoT experts are pushing towards integrating EC and IoT paradigms in order to support the critical IoT applications that require enhanced QoS (see Fig. (1)).

Fig. (3) shows the standard three-layer architecture of EC-assisted IoT paradigm. It is composed of the same layers of the conventional EC structure, where all the IoT "things" (i.e., devices and sensors) are considered as end users for EC. For the conventional IoT architecture, the EC layer does not exist. For the conventional EC architecture, there is an additional intermediate layer called "Core Networks" between Cloud layer and EC layer. There is also fog computing (FC) architecture, which is a standard that enables bringing cloud computing capabilities to the network edge. Although there is a tight overlap between EC and FC architectures [43], FC focuses more on the network infrastructure layer, while EC focuses more on the things layer [44]. To be more specific, FC enables repeatable structure in the EC concept, such that network developers can push computation capabilities out of the cloud computing layer to the EC servers in order to enable a robust and scalable performance. Whereas, EC assigns computation and processing resources from the cloud to the data-originating IoT devices at the network edge [45]. Another difference is that FC typically uses open standard technologies, whereas EC can use both open and proprietary technologies. It is noteworthy that this paper is dedicated to surveying security and privacy for the EC-assisted IoT paradigm. However, the interested reader can refer to [19], [40]–[42], [46]–[53] for related literature on the FC-assisted IoT paradigm.

In order to support the innovative IoT applications for the edge devices, and to enable the promising vision of the EC-assisted IoT paradigm, research community and industry have proposed a wide variety of EC architectures and technologies. Such technologies include the cloudlets mini servers [54]–[60], vehicular (or portable) edge computing (VEC) [61]–[66], and edge-cloud [67]–[70]. These technologies mainly enable the deployment of applications in harsh and rapid time-varying environments. There are also mobile edge computing (MEC) [71]–[76] and mobile cloud computing (MCC) [77]–[80] technologies, which enable the deployment of extensive-computation applications on the local IoT smart devices. This is by offloading a large portion of the applications locally on the devices themselves. Our main focus in this paper is on security and privacy issues in the EC paradigm in general.

Generally speaking, IoT can utilize the resources of both EC and cloud computing, such as the high computational capacity, large storage, and huge power capabilities. However, EC is more beneficial for time-sensitive applications that require fast response time with tolerable computational capacity and moderate storage space. On the other side, EC will benefit from IoT as well, by making IoT devices that have tolerable computation capacity act as EC nodes to provide services. Indeed, the explosive increase in the number and types of IoT smart devices will further push towards merging EC and IoT.

Although there is extensive research on conventional IoT cloud computing [11], [12], there are also several research works that investigate the feasibility of exploiting EC to assist IoT. The authors in [35] conduct a survey to analyze how EC can assist the performance of IoT networks. The performance of EC and cloud computing architectures are also compared in some IoT applications, such as smart transportation, smart city, and smart grid. In [37], the authors survey multi-access EC, and they present a holistic overview of this paradigm in relation with IoT. The integration of multi-access EC into IoT applications and their synergies are also analyzed and discussed. In addition, the technical aspects of this paradigm are also investigated to provide insight into different integration technologies in IoT multi-access EC. Ni *et al.* [36] examine the architecture of mobile EC and they discuss the potentials and advantages of using it to improve data analysis and computational efficiency for various IoT applications. The work in [38] investigates the key rationale, efforts, key enabling technologies, and typical applications of EC-assisted IoT. In [41], the authors present a survey on EC-assisted IoT literature in the period 2008–2018 including services, enabling technologies, and some open research directions. Caprolu *et al.* [81] discuss some of the technologies, scenarios, issues, and benefits of EC-assisted IoT. The authors in [44] present several case studies for EC-assisted IoT, such as cloud offloading and smart city/home, and they introduce several challenges and future research directions. The concept of industrial IoT (IIoT) is introduced in [82], in which the authors present the research progress and future architecture of EC-assisted IIoT. The authors also survey some research efforts related to security, task scheduling, routing, standardization, and data storage and analytics in the context of EC-assisted IIoT.

### A. Advantages of EC-Assisted IoT

There are several prominent advantages of integrating EC to assist the IoT, which can be classified into three main categories.

**1) Communication:** EC-assisted IoT networks have enhanced network performance in terms of reduced latency (both communication and computation), reduced bandwidth usage, reduced device power consumption, and reduced packet data overhead [35], [41]. Hence, the overall network performance in terms of communication is tremendously improved, which enables them to fulfill the QoS requirements of the time-sensitive IoT applications and services.

**2) Computation:** in EC-assisted IoT networks, data processing and computation will be offloaded to the edge servers, which relieves a massive burden from the centralized cloud servers. This guarantees enhanced network efficiency in terms of resource utilization and priority management [38].

**3) Storage:** since IoT end devices usually have limited storage capabilities, EC servers provide storage services to such devices. This is by migrating all the data generated or collected by the devices to storage servers. Doing so will assist in managing load balancing and failure recovery issues, leading to a significant enhancement in the QoS.

### B. Challenges of EC-assisted IoT

Although there are several advantages of utilizing EC-assisted IoT architecture, there still many key challenges encountered.

**1) Security and Privacy:** EC will encounter new and unforeseen security and privacy issues. IoT functionality requires the migration of services between local and global scales, which renders the network more vulnerable to potential malicious activities. In addition, since the users' privacy-sensitive information will be shared and/or stored at the EC servers, security and privacy become crucial challenges in such a distributed structure. This renders the EC-assisted IoT networks more vulnerable to cyber attacks and threats. Generally speaking, malicious attacks can be encountered during the three main processes of EC servers; communication, computation, and storage [5], [6], [10], [13], [18], [36], [39], [40], [42], [44]. Later in Sections III, IV, and V, we will provide a comprehensive analysis of the "security and privacy" issues.

**2) Network Heterogeneity:** EC-assisted IoT networks are heterogeneous, as they ensemble various network topologies, physical platforms, and servers. Hence, ensuring seamless operations for IoT devices in such a complex and sophisticated environment represents also one of the main challenges. For example, it would be quite challenging to program and control resources in applications running on different scattered and heterogeneous physical platforms.

**3) Resource Management:** controlling, managing, and optimizing the three main resources (communication, computation, and storage) of the decentralized EC-assisted IoT networks is also one of the crucial issues that must be properly investigated and addressed. This issue emerges due to the tremendous heterogeneity of service providers, IoT edge devices, applications, etc.

**4) Smart System Support:** the merging of smart IoT devices, such as meters, sensors, and actuators, will provide unprecedented opportunities for data collecting/sharing, resource allocation and optimization, and system management. Nonetheless, the challenge remains in how to enable multiple EC servers/nodes to store, process and share the collected data traffic from these multi-platform devices spanning wide geographical areas, in a manner that ensures optimal and timely management decisions.

### C. Definition of Security and Privacy in the Context of EC-Assisted IoT

As we mentioned previously, EC-assisted IoT systems manage a massive amount of information at the edge of IoT networks. Such information belongs to a range of low to high-sensitive applications and services of various vertical IoT sectors (see Fig. (1)). In the conventional implementations of EC-assisted IoT systems (see Fig. (3)), the data communication between end devices and EC nodes is accomplished through wired and/or wireless links. Whereas, the data communication from EC nodes to the cloud system utilizes either public or private networks [13]. Unfortunately, none of these EC-assisted system implementations is well-secured, making them vulnerable to huge security and privacy threats and attacks.

Several research works have analyzed such threats. The authors in [36] study the security, privacy, and some efficiency challenges of data processing in mobile EC. The opportunities for improving data security and privacy as well as enhancing computational efficiency with the assistance of EC, are also discussed. Solutions presented in their paper include secure data duplication and aggregation as well as secure computational offloading. In [40], the security mechanisms, threats, and challenges of some EC paradigms are analyzed. In [6], the authors describe the possibility of utilizing the attractive features and advantages of EC paradigms in enhancing some critical security and privacy issues in vehicular networks, particularly in revocation and authentication issues. The concept and features of EC-assisted IoT are introduced in [39], along with the requirements for its secure data analytics. The authors also analyze some prospective security and privacy threats and attacks, and they discuss some mechanisms for outsourcing data analytics. The authors in [5] provide an analysis of some of the data security and privacy attacks, and they describe countermeasure technologies in EC-assisted IoT networks.

Multi-access EC is a new paradigm that works as a complement for the centralized cloud architecture. It provides additional computing and storage resources at the edge of radio access networks and IoT applications. The authors in [13] conduct a survey to study the security challenges in mobile EC networks. The study focuses on security issues in systems of environment perception industrial IoT networks and mobile IoT based on a network of unmanned aerial vehicles (UAVs). The wireless medium is more vulnerable to attacks since it can be accessed by both authorized users and adversaries. Therefore, their study aims to discuss the security issues of the two aforementioned applications which exploit the benefits of mobile EC.

Creating a secured and privacy-preserving EC-assisted IoT ecosystem demands the implementation of different types of security and privacy mechanisms, requirements, and solutions. Section III explains the main security threats and attacks. Whereas, Section IV discusses the corresponding countermeasures, along with the related research work. Section IV also explains the main security/privacy mechanisms, and classify the related work based on security functions and services.

## III. Classifications Based on Security and Privacy Attacks and Threats

This section describes the key possible security and privacy attacks, their types, and their sources at different levels and layers (e.g., EC devices, communication and EC servers/nodes, and cloud servers) of EC-assisted IoT networks. Related research papers that survey each type will be also cited in each category.

*1) Malicious Hardware/Software Injection:* attackers can add unauthorized software/hardware components to the communication or EC node levels, that inject malicious inputs into the EC servers. This will enable adversaries to exploit service providers to perform hacking processes on their behalf, such as bypassing authentication, stealing data, reporting false data, or exposing database integrity [1], [10], [11], [39], [40]. Hardware injection attacks have several classifications, including 1) Node Replication, in which adversaries will inject a new malicious EC node to the network and assign it an ID number that is a replica of existing authorized node. Doing so will enable attackers to corrupt, steal, or misdirect data packets arriving at the malicious replica. In addition, node replicas can also even revoke legitimate EC nodes by implementing node-revocation protocols [1]. 2) Hardware Trojan, which is illegitimate access to integrated circuits (ICs), that makes attackers control the circuit and access data or even software running on these ICs. Trojans have two types; a) internally-activated Trojans, which can be triggered and activated if a particular condition is satisfied inside the ICs, and b) externally-activated Trojans, which are activated by sensors or antennas that interact with the outside world [1], [11]. 3) Camouflage, in which attackers inject a counterfeit EC node to the network, which will work as a normal EC node to generate, share, receive, store, process, redirect, or transmit data packets [1]. 4) Corrupted or Malicious EC Nodes, which are used to gain unauthorized access and control on the network, then injecting misleading data packets or even blocking the delivery of legitimate and true data packets [1], [10], [36], [37].

*2) Jamming Attacks:* in which attackers intentionally flood the network with counterfeit messages to exhaust communication, computing, or/and storage resources. This will render authorized users unable to use the infrastructure of the EC-assisted IoT network [39].

*3) Distributed Denial of Service (DDoS) Attacks:* outage attacks, sleep deprivation, and battery draining are the most famous types of DDoS attacks against EC nodes. In outage attacks, EC nodes stop performing their normal operations as they have been exposed to unauthorized access. In sleep deprivation, adversaries overwhelm EC nodes with an undesired set of legitimate requests. Such an attack is much harder to be detected. In battery draining, the battery of EC nodes or sensors/devices is depleted, so nodes failure or outage occurs. On the communication level, however, the most common DDoS attack is jamming the transmission of signals, which includes: 1) continuous jamming over all transmissions and 2) intermittent jamming by sending/receiving packets periodically by EC nodes [1], [2], [5], [10], [11], [13], [36], [37], [39]–[42].

*4) Physical Attacks or Tampering:* this attack happens if attackers can access the EC nodes/devices physically. In such a case, valuable and sensitive cryptographic information can be extracted, the circuit can be tampered with, and the software/operating-systems can be modified or changed [1], [10], [11], [13], [39]–[41].

*5) Eavesdropping or Sniffing:* adversaries covertly listen to private conversations, such as usernames, passwords, etc., over communication links. If sniffed packets contain session or control information of the EC nodes, such as nodes' configuration, nodes' identifiers, and password of the shared network, attackers can gain crucial information about the network [1], [10], [11], [39].

*6) Non-Network Side-Channel Attacks:* even if EC nodes are not transmitting any data, they may reveal critical information. For instance, the detection of known electromagnetic/acoustic signals or protocols from medical devices can lead to serious privacy issues, as critical information about the patient and device can be leaked [1], [11].

*7) Routing Information Attacks:* attackers alter routing information by redirecting or dropping data packets at the communication level. The malicious EC nodes might be: 1) Black Holes, which drain all network's packets, 2) Gray Holes, which drain selective packets, 3) Worm Holes, in which attackers will first record packets at one network location then migrate them to another location, or 4) Hello Flood, in which a high-power malicious EC node broadcasts 'HELLO PACKETS' to all nodes claiming to be their neighbor [1], [11], [39].

*8) Forgery Attacks:* in which attackers inject new fraudulent data packets and interfere with the receiver causing system damage or failure. These data packets are inserted to communication links using methods such as 1) inserting malicious data packets that seem legitimate, 2) capturing then modifying data packets, and 3) replication of previously exchanged packets between two EC nodes/devices [1], [11], [13], [39], [40].

*9) Unauthorized Control Access:* neighboring EC nodes communicate with each other to access or share their data. However, if attackers can access one of the unsecured EC nodes, it is possible to control the whole neighboring nodes [1], [11].

*10) Integrity Attacks Against Machine Learning:* machine learning methods used in EC-assisted IoT are also vulnerable to two types of attacks; 1) causative, in which attackers change the training process of machine learning models by manipulating and injecting misleading training dataset, and 2) exploratory, in which attackers utilize vulnerabilities without changing the training process [1].

*11) Replay Attack or Freshness Attacks:* in which attackers capture and record data traffic for a particular period of time and then use this historical data to replace the current real-time data. Doing so will cause energy and bandwidth consumption of EC nodes as well as other adverse effects [11], [13].

*12) Inessential Logging Attacks:* if log files are not encrypted, this type of attacks can lead to damage in EC-assisted IoT systems. Therefore, system and infrastructure developers must log events, such as application errors and attempts of unsuccessful/successful authorization/authentication [40].

*13) Security Threats from/on IoT Devices:* cyber attacks on EC devices include mobile Botnets, ransomware, and IoT malware. In 2017, over 1.5 million attacks originated from mobile malware were reported [36]. Such threats bring security concerns towards both edge users and applications leading to data leakage/corruption or even application death [36], [39].

*14) Privacy Leakage:* EC nodes' functionalities may need to extract personal information from the data generated by user devices. Some might be sensitive, e.g. personal activities, preferences, and health status; however, others might not be, e.g. air pollution index, public information, and social events. Nonetheless, all information must belong to data owners. Unfortunately, they could be shared with other users or network entities without granting permission from the information owners, which makes them vulnerable to intruders during data transmission/sharing. Attackers can exploit the location awareness of EC nodes (e.g., Wi-Fi hotspots and base stations (BSs)) to detect and track the device's physical position or other sensitive information from the physical location of these EC nodes. Moreover, if user devices establish connections to multiple EC nodes simultaneously in order to access a particular service, the physical location might be precisely detected using positioning techniques [36], [39], [40].

*15) Other Attacks:* EC-assisted IoT paradigm is a combination of heterogeneous resources and devices manufactured by various vendors. Since there is neither a generally-agreed framework nor standard policies for the implementations of this paradigm, there still many security and privacy threats undetected.

## IV. CLASSIFICATIONS BASED ON SECURITY AND PRIVACY COUNTERMEASURES AND FUNCTIONS

This section explains the main strategies and solutions developed to countermeasure the security and privacy attacks and threats explained in the previous section. In addition, classification based on security functions and services is also provided.

### A. Classifications of Solutions and Countermeasures

*1) Countermeasures for Malicious Hardware/Software Injection:* there are several effective techniques developed to tackle this; 1) Side-Channel Signal Analysis, which is used to detect both: a) hardware Trojans, by implementing timing, power, and spatial temperature testing analysis and b) malicious firmware/software installed on IoT EC nodes/devices, by detecting unusual behaviors of nodes/devices, e.g., a significant increase in their heat, execution time, or power consumption [1]. 2) Trojan Activation Methods, which are used to compare the outputs, behavior, and side-channel leakages of Trojan-inserted vs Trojan-free circuits, in order to detect and model malicious attacks [1], [39]. 3) Circuit Modification or Replacing, this is also an effective countermeasure against physical/hardware, Trojan, and side-channel attacks. This countermeasure includes: a) tamper-preventing and/or self-destruction, in which EC nodes are physically embedded with hardware to prevent malicious attacks, or in the worst cases the EC nodes destruct themselves and/or erase their data, b) minimizing information leakage, by intentionally adding random noise or delay to the data, implementing a constant execution path code, and balancing Hamming weights, and c) embedding Physically Unclonable Function (PUF) into the circuit hardware, which enables device identification and authentication to detect Trojan activities [1].

*2) Policy-Based Mechanisms:* which are used to detect any violation of policies, by ensuring that standard rules are not breached. For example, they detect any abnormal requests to the EC node that try to cause sleep deprivation or battery-draining [1].

*3) Securing Firmware Update:* the update of the network's firmware can be reliably established either remotely (e.g., EC servers broadcast messages to announce and share the updated version of firmware) or directly (e.g., using USB cables). Both methods require authentication and integrity to ensure security updates [1].

*4) Reliable Routing Protocols:* in which EC nodes create a table of trusted nodes for sharing sensitive and private information. Further explanation of this type of countermeasures can be found in [1], [10], [11] and the references therein.

*5) Intrusion Detection System (IDS):* which is the second line of defense employed to mitigate security threats by 1) monitoring network's operations and communication links, 2) reporting suspicious activities, such as when predefined policies are breached or when invalid information is injected into the system, and 3) detecting routing attacks (e.g., spoofing or modification of information) as well as Black Hole attacks [1], [11], [35], [40]. The authors in [8] propose an IDS architecture for EC-assisted IoT, which integrates a trust evaluation mechanism and service template with balanced dynamics. In their proposed solution, the EC network is designed to minimize resource consumption, whereas the EC platform is designed to ensure the extensibility of the trust evaluation mechanism. Lin *et al.* [15] propose a general EC IDS architecture, which shows an efficient fair resource allocation in EC-assisted IoT systems.

*6) Cryptographic Schemes:* which are strong and efficient encryption countermeasure strategies utilized to secure communication protocols against various attacks, such as eavesdropping and routing attacks. Although there is a wide variety of encryption/decryption strategies developed to enhance network security and privacy, such solutions are applicable for wired networks. Unfortunately, EC nodes are typically tiny sensors with limited resources, e.g., battery power, computing/processing capabilities, and storage memory. Therefore, employing standard encryption/decryption techniques will increase memory usage, delay, and power consumption [1], [10], [11], [39]. The authors in [5] thoroughly explore the architectures and ideas of several key crypto-systems, such as proxy re-encryption, attribute-based encryption, searchable encryption, identity-based encryption, and homomorphic encryption. Chen *et al.* [17] propose a non-cryptographic security access method for the EC-assisted IoT paradigm. Unlike the conventional cryptographic algorithmic-based security access scheme, their proposed solution does not require password authentication, as it mainly relies on the differences in the hardware of the heterogeneous wireless access devices. The work in [18] proposes a secure data-sharing scheme for EC-assisted IoT smart devices. The proposed scheme uses both public and secret key encryptions. In addition, a searching strategy is also presented that enables authorized users to perform secure data search within shared, encrypted, and stored data in EC-assisted IoT networks, without leaking data, secret key, or keyword. In [12], the authors present an architecture based on data proxy concept, which applies process knowledge in order to enable security via abstraction as well as privacy via remote data fusion.

*7) De-patterning Data Transmissions:* this strategy prevents side-channel attacks, by intentionally inserting fake packets that change the traffic pattern [1], [5], [39].

*8) Decentralization:* this strategy ensures anonymity, by distributing the sensitive information through EC nodes such that no node has complete knowledge of the information [39].

*9) Authorization:* this strategy prevents responses to requests originated by attackers or malicious EC nodes. It scrutinizes if an entity (e.g., service provider, EC node/device, router, etc.) can access, control, modify, or share the data [1], [5], [11], [42].

*10) Information Flooding:* this strategy prevents intruders from detecting and tracking the location of the information source [10].

*11) Prior Testing:* in which a behavioral test of the components of EC-assisted IoT network (EC routers/nodes, servers, etc.) is conducted prior to the actual operation. This is accomplished by applying special inputs, pilot, and/or token signals to the network and monitoring their outputs. This solution mainly aims at identifying the possible attacks, simulate them, and evaluate their impacts on the EC-assisted IoT paradigm. It also classifies the information to define which must be logged and which is sensitive to be shared or stored [1], [39].

*12) Outlier Detection:* attacks against machine learning methods aim at injecting data outliers to the training dataset. Such attacks are drastically mitigated using statistical data analytics methods [10], [39].

*13) Secure Data Aggregation:* which is a highly-secure, privacy-preserving, and efficient data compression strategy. In this scheme, individual devices will use homomorphic encryption scheme (such as Brakerski-Gentry-Vaikuntanathan (BGV) cryptosystem) to independently encrypt their own data, and then sends it to the EC nodes. The later will aggregate all data in order to compute the multiplication of individual data, and then send the aggregated results to the central cloud servers [10], [36], [39].

*14) Secure Data Deduplication:* removing data redundancy and utilizing the bandwidth in IoT networks require to remove the replicate copies of data on intermediate EC nodes. Unfortunately, this will render sensitive information disclosed to intruders. To countermeasure this threat, secure data deduplication is used, in which intermediaries are allowed to access the replicated data without gaining any knowledge about it [36], [39].

*15) Secure Data Analysis:* the explosive advances in EC devices have enabled the shift of some artificial intelligence (AI) functionalities from the centralized cloud to EC devices/nodes. This will improve security, privacy, and latency. For example, partitioning network functionality execution among EC nodes/devices and the central cloud enables individual nodes/devices to locally and independently train their own models and then only share their individual trained models rather than their respective private training dataset [36], [39].

*16) Authentication:* in the EC-assisted complex environment, it is required to make entities mutually authenticate one another across different trust domains. This includes single/cross-domain and handover authentication. Such schemes are discussed in detail in [5], [10], [11], [39], [40], [42].

*17) Combining EC and Blockchain Technologies:* blockchain is an emerging strategy that provides a trusted, reliable, and secure foundation for information transactions and data regulation between various operating network edge entities. It creates rules that enable decentralized systems to jointly perform decisions about the execution of particular transactions, depending on voting and consensus algorithms. This will; 1) ensure a secure audit-level tracking of EC-assisted IoT data transactions and 2) eliminate the requirement for a central trusted intermediary between the communicating IoT edge devices [41]. The authors in [14] develop a secure and distributed data storage and sharing scheme for vehicular EC networks based on integrating the smart contract technologies with consortium blockchain. Gai *et al.* [16] combine EC and blockchain technologies and they propose a permissioned blockchain EC model that addresses the privacy-preserving and energy security of smart grid EC-assisted IoT networks. They also present a security-aware strategy based on smart contracts running on the blockchain, and they evaluated the efficiency of their proposed scheme experimentally.

Table (I) provides a summary list of the papers discussed in this section. They are classified based on security attacks and threats as well as based on solutions and countermeasures they discussed. It is noteworthy that although some of the security and privacy-related concepts, attacks, and solutions presented in the original papers were in the context of conventional centralized cloud-based IoT, some are also applicable or can be extended to the EC-assisted IoT paradigm as well.

Table I: List of the Papers Discussed in Section (IV), Classified Based on Security and Privacy Attacks and Countermeasures They Provide.

| Ref. | | [35] | [10] | [13] | [37] | [14] | [8] | [11] | [36] | [1] | [6] | [2] | [15] | [16] | [17] | [38] | [39] | [40] | [12] | [41] | [5] | [18] | [42] |
|---|---|---|---|---|---|---|---|---|---|---|---|---|---|---|---|---|---|---|---|---|---|---|---|
| Attacks and Threats | 1 | ✓ | | | ✓ | | | ✓ | ✓ | ✓ | | | | | | | ✓ | ✓ | | | | | ✓ |
| | 2 | | | | | | | | | | | | | | | | ✓ | | | | | | |
| | 3 | | ✓ | ✓ | ✓ | | | ✓ | ✓ | ✓ | | ✓ | | | | | ✓ | ✓ | | ✓ | ✓ | | ✓ |
| | 4 | | ✓ | ✓ | | | | ✓ | | | | | | | | | ✓ | ✓ | | ✓ | | | |
| | 5 | ✓ | | | | | | ✓ | | ✓ | | | | | | | ✓ | | | | | | |
| | 6 | | | | | | | ✓ | | ✓ | | | | | | | | | | | | | |
| | 7 | | | | | | | ✓ | | ✓ | | | | | | | ✓ | | | | | | |
| | 8 | | | ✓ | | | | ✓ | | ✓ | | | | | | | ✓ | ✓ | | | | | |
| | 9 | | | | | | | ✓ | | ✓ | | | | | | | | | | | | | |
| | 10 | | | | | | | | | ✓ | | | | | | | | | | | | | |
| | 11 | | | ✓ | | | | ✓ | | | | | | | | | | | | | | | |
| | 12 | | | | | | | | | | | | | | | | | ✓ | | | | | |
| | 13 | | | | | | | | ✓ | | | | | | | | ✓ | | | | | | |
| | 14 | | | | | | | ✓ | | | | | | | | | ✓ | ✓ | | | | | |
| Solutions and Countermeasures | 1 | | | | | | | | | ✓ | | | | | | | ✓ | | | | | | |
| | 2 | | | | | | | | | ✓ | | | | | | | | | | | | | |
| | 3 | | | | | | | | | ✓ | | | | | | | | | | | | | |
| | 4 | | ✓ | | | | | ✓ | | ✓ | | | | | | | | | | | | | |
| | 5 | ✓ | | | | ✓ | ✓ | ✓ | | ✓ | | ✓ | | | | | ✓ | | | | | | |
| | 6 | | ✓ | | | | | ✓ | | ✓ | | | | | ✓ | | ✓ | | ✓ | | ✓ | ✓ | |
| | 7 | | | | | | | | | ✓ | | | | | | | ✓ | | | ✓ | | | |
| | 8 | | | | | | | | | | | | | | | | ✓ | | | | | | |
| | 9 | | | | | | | ✓ | | ✓ | | | | | | | | | | ✓ | | | ✓ |
| | 10 | | ✓ | | | | | | | | | | | | | | | | | | | | |
| | 11 | | | | | | | | | ✓ | | | | | | | ✓ | | | | | | |
| | 12 | | ✓ | | | | | | | | | | | | | | ✓ | | | | | | |
| | 13 | | ✓ | | | | | | ✓ | | | | | | | | ✓ | | | | | | |
| | 14 | | | | | | | | ✓ | | | | | | | | ✓ | | | | | | |
| | 15 | | | | | | | | ✓ | | | | | | | | ✓ | | | | | | |
| | 16 | | ✓ | | | | | ✓ | | | | | | | | | ✓ | ✓ | | | ✓ | | ✓ |
| | 17 | | | | ✓ | | | | | | | | | | ✓ | | | | | ✓ | | | |

## B. Classification Based on Security Functions and Services

Security is one of the main concerns in EC-assisted IoT systems. Due to the diverse enabling technologies which constitute IoT networks, several security mechanisms need to be employed to support security. EC-assisted networks are usually comprised of a combination of virtualization platforms, wireless networks, peer-to-peer, and distributed systems. It is considered as a big concern, not only to provide protection to all these varied components but also to enable these diverse security mechanisms to coordinate and cooperate. Security and privacy objective is to ensure confidentiality, integrity, and availability of the system and its assets [19].

Furthermore, storing the data collected by IoT devices at the edge nodes might create a privacy issue as these edge devices are more vulnerable to attacks than centric cloud servers [35]. Therefore, privacy protection is a major issue in EC, and hence effective mechanisms should be developed to preserve the privacy of users in the EC-assisted IoT environment. Security and privacy objectives can be met by developing different protection mechanisms for authentication, access control, data transmission, storage, and computation. Each one of these security functions has several issues. In the following sections, we analyze and classify them based on their impact on security and privacy objectives and EC IoT services.

### a. Authentication

One of the main security aspects of EC-assisted IoT paradigms is authentication. Edge networks are composed of multiple distributed entities that coexist and interact within ecosystem domains. Hence end users, edge devices, service providers, and data centers need to authenticate each other, which represents a challenge that requires a sophisticated multilevel authentication mechanism. It is not only necessary to assign an identity to every entity in the domain, but also all the entities need to authenticate each other mutually. These authentication issues demand complex authentication controls to prevent external adversaries from attacking system assets and resources [40]. The following subsections discuss some issues related to authentication mechanisms for the EC-assisted IoT networks.

#### 1) Identity Management and Key Exchange for Multiple Distributed Entities: given the limited resources of IoT devices, inter-realm authentication systems, and identity federation mechanisms are two of the solutions that can be explored in this context. Besides the cooperation feature of these mechanisms, they allow devices and users to provide proof of their identity without a central authentication server. Applicability of distributed authentication mechanisms is still an issue in EC-assisted IoT paradigms as in some cases, central authentication is still necessary to manage the identities of parts of the infrastructure [40]. Esiner and Datta [28] propose a layered security mechanism for EC-assisted IoT

networks based on a distributed multi-factor authentication without third-party interference. It mainly depends on knowledge and possession factors to prove user identity. Data are distributed among several data storage centers and retrieved based on a password of the user's selection from multiple other passwords corresponding to each server. However, due to the decentralized design of this protocol, users will not be able to restore their data if they forget the initial password.

The proposed authentication mechanism from Jan *et al.* [20] depends on sharing a session key between nodes and their cluster head. To identify authorized nodes for the cluster heads, base stations (BSs) receive requests from edge nodes for authorization. The proposed authentication mechanism involves different levels of identity definition and authentication between edge cluster heads, edge devices, and BSs. The issue of the distribution and management of the encryption keys was indicated as part of the future work of King *et al.* [83]. They propose a two-phases transmission security mechanism where one layer represents the connection between IoT constrained resources devices and the edge device. Whereas, the other layer secures the transmission between the gateway and the end server.

Though end devices need to have a single authentic identity and secure key, applications within each device may require additional key exchange mechanism for further application-related security. Some studies indicated and explored the complexity of identity assigning and key management of cross-application mechanisms. The authors in [25] indicate that user devices might be engaged in multiple applications and require multiple security keys, which may increase security risk and key disclosure. The proposed solution to this problem as indicated in their study suggests that each IoT domain generates and maintains security keys for IoT devices belong to each domain. Each device has to maintain a set of security keys for each application. This may result in a big number of keys and will increase the complexity of key management. To solve this problem, the study adopted hierarchy-based key management, where services and applications credentials are composed of multiple keys based on the level of the application in the hierarchical schema.

*2) Development of Resources Efficient Authentication Mechanisms:* intruders usually aim to access the network and perform malicious actions, such as misleading data injection or malicious code injection. To prevent intruders from accessing the network, a sophisticated but efficient authentication mechanism is required. In EC-assisted IoT, some edge devices have limited resources, and hence traditional complex authentication mechanisms might not be applicable [21]. Therefore, developing an authentication mechanism that utilizes the available resources efficiently is an issue. In [21], the authors propose an efficient Edge-Fog authentication scheme, to securely allow mutual authentication between Fog user and any Fog server. The proposed scheme does not depend on public key infrastructure (PKI) to perform the authentication but forces Fog users to store only one long-lived master secret key which will allow to mutually authenticate with any Fog server in the domain. To alleviate the problem of constrained resources of some edge devices, Sha *et al.* [23] suggest moving security functions such as authentication to devices that have enough resources to handle the computation need of other edge devices. They develop a comprehensive architecture composed of several modules, each of which is responsible for handling a certain security service as a response to different challenges of EC-assisted IoT. The security analysis module is responsible for assigning security functions to edge devices based on information about them collected by another module.

*3) Maintaining Authentication Sessions:* initiating as well as maintaining authentication sessions of edge users is a general security issue in EC-assisted IoT systems. Using only username/password to authenticate users might not be secure enough, the authors in [24], therefore, suggest a multi-factor authentication mechanism. Their proposed solution maintains the session state through real-time identity monitoring. Edge devices keep updating the state of the connection with the authenticated user by regularly requesting additional authentication methods such as collecting information about the normal behavior of the user or matching the current state of the user with valid former states. If a deviated or abnormal behavior is detected a request for re-verification is triggered.

*b. Access Control*

For any two entities in a system to share resources, they essentially need to have credentials and access policies. Most of the operations in EC-assisted IoT networks include requesting to access resources, sending or receiving data, and performing processing. If there is no defined authorization mechanism, access to system resources will have no restrictions, and hence illegal operations on IoT devices can be launched. To develop an EC-assisted IoT authorization infrastructure, it is crucial to enforce security access policy in each trust domain. Entities within the trust domain should be able to identify and verify each others' identities. They also need to define the level of resource allocation [5]. The following subsections indicate some authentication and access control related issues.

*1) Detection and Management of Transitive Access Control:* one of the access control issues is the transitive access between edge devices or entities in EC-assisted IoT networks. Granting access to a certain device to access resources through another intermediate device should be controlled, as this may expose resources to malicious or unauthorized access. Sha *et al.* [23] propose a security analysis module to detect transitive access and judge whether it is legal or not. The detection mechanism is developed based on a representation of access requests as a directed graph.

*2) Control Access to Fine-Grained Edge Node Components and Services:* each edge device hosts multiple applications and services. Controlling access to each element of these services and applications represents a challenge. Edge devices have to grant access to resources based on a predefined authorization policy. Maintaining and forcing access policy may consume resources and thus, an efficient and secure mechanism to maintain and force this policy is required. The authors in [25] propose a fine-grained access control based on the keys and attributes of edge users and IoT devices. This may allow for adopting different security measures by considering each security service as an object. The attribute-based encryption mechanism combines the verification of the IoT device key with its attributes in addition to the access policy to encrypt messages and hence, only authorized edge devices and users can have access to these messages. The attribute-based access control is introduced in EC-assisted IoT networks to reduce the number of rules resulting from role explosion. It protects data security by sharing data between multiple users. As indicated by Cui *et al.* [26], the attribute-based encryption provides scalable fine-grained access control over IoT edge resources and data. The authors in [26] adopt this mechanism into the EC-assisted IoT paradigm through the establishment of third-party key distribution and the availability of a secure channel.

*3) Supporting Access Control for Dynamic Scalable IoT Networks:* most of the EC-assisted IoT networks have a dynamically evolving architecture in terms of the number of devices, services, and users. Providing access control strategy that meets the growing requirements of these networks is a challenging problem. Maintaining access control based on static constant features of objects and entities may become obsolete by time and make the system vulnerable to various attacks. Some solutions in the literature [25]–[27] propose a scalable access control mechanism based on different dynamic properties. The solution proposed in [27] uses a capability-based access schema. They argue that the attribute-based encryption mechanism used in [25], [26] may not meet the requirements of EC-assisted IoT networks as it might increase effort and complexity of policy management as the number of devices increases and size of the network expands, which may not make it a perfect solution for the scalable distributed EC-assisted IoT networks. The authors in [28] propose a layered security mechanism based on a distributed multi-factor access control. Their proposed protocol does not require a third-party interference, and it mainly depends on knowledge and possession factors to prove user identity. By distributing data among several edge data storage centers and derive several passwords for storage servers based on an initial seed password, they provided a decentralized access control mechanism suitable for scalable dynamic EC-assisted IoT networks.

*c. Data Security*

Since data is the main element of IoT systems, it needs to be protected during transmission, computation, and storage. The development of the EC-assisted IoT paradigm aimed basically to alleviate latency and reduce data transfer between cloud servers and IoT edge devices. Reducing the amount of data transmission between network devices will decrease the exposure of these data to attacks. Therefore, the EC-assisted IoT paradigm provides a more secure architecture than the other computing paradigms, such as cloud computing. In EC-assisted IoT, the edge nodes are responsible for carrying a significant part of processing tasks by receiving input from other edge nodes and sending output to end-users or cloud servers. Hence some input and output data transmission over the network is still exposed and needs protection. Moreover, the data is stored at the edge devices and thus, a secure mechanism is required for storage protection. Some issues associated with data storage and transmission will be discussed in the following subsections.

*1) Data Storage Auditing and Encryption Latency:* one of the main similarities between cloud computing and EC is data outsourcing. Data is usually stored in edge servers, and hence there is a possibility of data loss, disclosure, or modification. Therefore, provision for data storage auditing is one of the most important solutions. Several services are provided by the infrastructure providers, including third-party auditing services, which are usually associated with a set of auditing policies. Several other techniques can be adopted to ensure confidentiality and integrity. Encryption is one of these methods that can also be utilized to check for the untrusted network. However, data auditing controls and data encryption mechanisms should be as efficient as possible, given that the main purpose of the EC-assisted IoT paradigm is to reduce latency and improve response time [5].

*2) Support Multiple Encryption Mechanisms:* providing security to real-time data transmission between edge devices represents another challenge. To secure data transmission over the EC-assisted IoT network, Jan *et al.* [20] propose an end-to-end encryption framework. Their proposed framework aims at providing security to real-time multimedia streams for smart cities. The edge IoT devices usually have different levels of computing and storage resources, hence different levels of encryption mechanisms are required to fit the capabilities of edge devices. Providing different types of encryption levels is a challenge, and allowing for interconnectivity between different transmission encryption mechanisms is also another challenge.

The framework proposed in [20] includes an authentication mechanism to initiate an encrypted data transmission using different levels of encryption complexity based on the type of the destination (edge node or cloud server). Sha *et al.* [23] develop a protocol mapping module to assign different transmission protocols to different edge devices based on their resources. With different transmission protocols, comes the problem of interconnectivity, which is caused mainly by the heterogeneity of communication protocols used by edge devices. The interface manager module proposed by [23], is designed to handle this issue by forwarding the package to the edge layer device that supports the detected communication protocol. Moreover, the authors in [83] propose a two-layer transmission security mechanism where one layer represents the connection between IoT constrained resources devices and the edge device (gateway). The transmission at this layer uses Advanced Encryption Standard (AES)-128 encryption standard. The other layer secures the transmission between the gateway and the end server. This layer uses Hypertext Transfer Protocol Secure (HTTPS) to secure the transmission between the gateway and the server. The study does not suggest any authentication mechanism, and it added data integrity and availability as future work. Furthermore, the distribution and management of encryption keys are found to be an issue that can be addressed as future work.

*3) Providing Protection to Distributed Decentralized Data Storage:* outsourcing data at the edge servers poses several security issues, for the decentralized distributed EC-assisted networks. One of the imposed issues is the capability to store data in a decentralized environment, where the network is rapidly growing and no central authentication or authorization mechanism is provided to secure access to this data. In [28], the authors propose a security layered mechanism for decentralized edge data storage. They established their solution for multi-factor access control. Data storing and retrieving can be established without the need for a third-party. It is mainly based on multi-factor several passwords as per the number of storage servers. The server password derived from an initial seed password.

*d. Computation*

One of the main objectives of security in EC-assisted IoT networks is to ensure the integrity and confidentiality of data computation. Data encryption is one of the security mechanisms, which can be employed to prevent data visibility or disclosure. Computation centers within the EC-assisted network have the provisions to offload some of the processing of the data to each other. Therefore, they need to verify the data generated by other computation centers and establish trust between the two data centers. Users also need to verify the validity and security of the acquired data. Other types of issues in security of EC include the development of security solutions on top of the EC distributed infrastructure. Due to the constrained resources and distributed, heterogeneous, and scalable architecture of EC-assisted IoT networks, deployment of security services and applications over these networks represents one of the main challenging problems [5]. The following subsections indicate two computational challenges of the development of security architectures.

*1) Distribution of Security Services and Functions:* edge devices vary in terms of resources, location, and availability. Identifying the best strategy to disseminate security functions and services over edge devices represents a challenge. Sha *et al.* [23] develop a security mechanism, which depends on distributing security services such as firewalls and intrusion detection over multiple edge devices, given the available resources of each device. The authors in [22] suggest that there are similarities between living organisms and IoT deployments in terms of security challenges. Therefore, they proposed a security architecture design similar to the virtual immune system for protecting the devices in the EC-assisted IoT. They defined cell components represented by software agents, which is responsible for monitoring, collecting information, and performing actions, whereas the kernel is responsible for making decisions and situation analysis based on information collected by the cells software agents. The main purpose of this design is to protect the EC-assisted IoT ecosystem from external intruders by monitoring traffic and data transmission in addition to other types of data collected from IoT devices.

*2) Flexibility to Support Various Security Protections for Diverse IoT Applications:* The development of security solutions for EC distributed heterogeneous architectures is a challenging problem. Some security solutions might not be applicable to all types of edge devices and applications, thus building a flexible security solution that does not require a fundamental change in the infrastructure of the different IoT networks is considered a security issue. The authors in [25] propose such a reconfigurable security framework for EC-assisted IoT networks.

*e. Trust Modeling*

The development of trust modeling for IoT in general and EC, in particular, is increasing. It is generally targeted to protect against internal attacks, where IoT devices are more vulnerable to internal intruders. External attacks usually mitigated using different types of controls such as authentication, encryption, and authorization. Protection against internal attacks, on the other hand, requires not only traditional security mechanisms but also other types of security controls such as trust modeling techniques. In many cases, the internal attacker employs some IoT devices in the network to initiate the attack. Therefore, maintaining a trust evaluation mechanism can be a solution to identify the source of the internal attack to contain, reduce or eliminate the threat [84]. The following subsections discuss some trust modeling issues related to the EC-assisted IoT networks.

*1) Maintain Trust for Dynamic Scalable Edge Networks:* in the EC-assisted IoT paradigm, the trust evaluation mechanism is moved from the cloud to the edge devices. In the trust evaluation mechanism proposed by Wang *et al.* [84], IoT edge devices can only perform simple direct trust estimations, and they forward exceptions and abnormal calculations to the edge servers for verification and management. Their proposed mechanism considers two modes of architecture, the fixed mode, and the moving mode. For the moving mode, the key issue is to develop a strategy to update the state of the trust of the moving IoT edge devices in the network. A hierarchical architecture was proposed by their study to alleviate the problem of the moving devices. Collecting trust information about IoT devices is accomplished at the edge devices, which performs state analysis and maintains the entire trust state of the EC-assisted IoT network. The study assumes the existence of an edge platform that is composed of powerful edge servers to perform complex operations such as service templates establishment. In [29], the authors propose a multi-weighted distributed reputation management framework for vehicular EC. To alleviate the problem of the scalability of vehicular networks, they employed several types of edge devices such as gateways and base stations (BSs) to collect and process trust information from vehicles. The data then forwarded to edge servers that communicate with each other and exchange information. However, Yuan *et al.* [30] provide a trust computing mechanism for which edge devices are responsible for not only collecting trust feedback from different sources but also performing the computation without relying on the central network. This distributed computing architecture provides support to the scalable EC-assisted IoT networks.

*2) Maintain Consistent Reliable Distributed Trust Information in Edge Devices:* in cloud computing, cloud servers are responsible for collecting information from IoT devices and performing computations. IoT devices are just responsible for sensing and reporting, controlling, etc. Establishing reliable and efficient trust management is performed by cloud servers. In the EC-assisted IoT paradigms, on the other hand, edge devices and edge servers share the responsibility of establishing and maintaining trust information about IoT edge devices, users, applications, etc. Sharing and processing trust information in a distributed manner raise several issues in terms of maintaining trust information in edge devices and servers as consistent as possible. The authors in [29] suggest moving trust information from edge devices to edge servers to maintain efficient and accurate multi-weighted updates of trust information in a timely manner. Yuan *et al.* [30] propose an adaptive algorithm to collect and maintain the overall trust of IoT edge devices, which depends on objective information based on entropy theory. The algorithm is proposed to maintain accurate and consistent evaluation of trust information.

Table II: List of the Papers Discussed in Section (III), Classified Based on Security Services and Functions.

| Ref. | Authentication | | | Access control | | | Data Security | | | Computation | | Trust | | Privacy | |
|---|---|---|---|---|---|---|---|---|---|---|---|---|---|---|---|
| | 1 | 2 | 3 | 1 | 2 | 3 | 1 | 2 | 3 | 1 | 2 | 1 | 2 | 1 | 2 |
| [40] | ✓ | | | | | | | | | | | | | | |
| [28] | ✓ | | | | ✓ | | | ✓ | | | | | | | |
| [20] | ✓ | | | | | | | ✓ | | | | | | | |
| [83] | ✓ | | | | | | | ✓ | | | | | | | |
| [25] | ✓ | | | | ✓ | ✓ | | | | | ✓ | | | | |
| [21] | | ✓ | | | | | | | | | | | | | |
| [23] | | ✓ | | ✓ | | | | ✓ | | ✓ | | | | | |
| [24] | | | ✓ | | | | | | | | | | | | |
| [26] | | | | | ✓ | ✓ | | | | | | | | | |
| [27] | | | | | | ✓ | | | | | | | | | |
| [5] | | | | | | | ✓ | | | | | | | | |
| [22] | | | | | | | | | | | ✓ | | | | |
| [84] | | | | | | | | | | | | ✓ | | | |
| [29] | | | | | | | | | | | | ✓ | ✓ | | |
| [30] | | | | | | | | | | | | ✓ | ✓ | | |
| [31] | | | | | | | | | | | | | | ✓ | |
| [32] | | | | | | | | | | | | | | | ✓ |
| [33] | | | | | | | | | | | | | | | ✓ |

*f. Privacy*

Moving data processing to edge devices raises an issue of preserving the privacy of user's data, behavior, and location. User data can be leaked, misused, or stolen which may discourage users from integrating EC-assisted IoT networks. Some curious adversaries who have the authority to access the data, such as service providers or edge data centers, might misuse or exploit personal data of users [5]. Moreover, edge devices are distributed and scattered in wide and open areas; therefore, the central controlling of these edge devices might be difficult. If one of the edge nodes compromised, intruders might use it as an entry point to the EC-assisted IoT network. The intruder exploits this vulnerability to steal users' personal information and private data that is exchanged between edge devices. The following subsections discuss privacy issues related to the user's identity, data, and location [42].

*1) Identity and Data Privacy:* generally, the privacy and security issues of EC-assisted IoT have recently gained the attention of the industry [31]. This is due to the fast-growing interest in these networks since they provide several advantages, including latency alleviation. Du *et al.* [31] confirmed that privacy issues analysis in EC has received little attention especially for data science and machine learning applications. Their study considers preserving the privacy of processing big data using machine learning. They mention that edge nodes are distributed randomly over the network which makes controlling them infeasible. If one of the nodes has poor security controls, it might become the fuse of the intruders malicious attack. To preserve the privacy in machine learning applications for EC-assisted IoT, they propose a machine learning privacy architecture for data aggregation and collection which consists of three levels. The system-level management, which is the core of the architecture. It is responsible for controlling the whole system and provides access to users and other parties. The second layer represents the host level virtualization layer of the proposed architecture. The last one is the network level layer, which preserves information collection at the network layer. Some machine learning EC solutions have been proposed to move processing to the edge device to maintain the privacy of the user's identity and data. Data transmission to the edge server or the cloud server is no longer required, and hence private information of users remains enclosed at the edge devices. The solution developed by [34] is proposed to anonymize the edge devices. The proposed application is crowd management (or crowd counting). Although they would process full RGB images and data at the edge, only aggregated counts would ever leave the edge, thereby effectively anonymizing any privacy-sensitive information, which was a very sensitive goal in the region. This computing mechanism is proposed to hide users's identities and can be considered a challenge and an opportunity.

*2) Location Privacy:* there are many web services and applications which provide location-based functions. Users need to submit their location to the service provider to have access to services. In many cases, location information leakage represents a definite danger and real concern to users. The authors in [32] introduce a system for mobile online social networks, which provides a flexible privacy-preserving location sharing. The system can identify untrusted strangers among social relations within a certain range. It hides location information by separating the storage of user identities and anonymous location information and then storing them in two separate entities. If one of the storage entities leaked or attacked, information about the location will be harmless because it will not reveal user identities. Chen *et al.* [33] propose a scheme to preserve location information of mobile users. The schema employs Markov Chain for distributed cache pushing proxies, which can divide location information into groups and store them separately. The location information is preserved by receiving location-based data from the cache proxies without revealing their real locations to service providers.

Table III: Relationship Between Security Countermeasures Discussed in Subsection (IV-A) and Security Functions Discussed in Subsection (IV-B)

| | ID | Authentication | Access Control | Data Security | Computation | Trust | Privacy |
|---|---|---|---|---|---|---|---|
| Solutions and Countermeasures | 1 | ✓ | ✓ | | | | |
| | 2 | ✓ | ✓ | | | ✓ | |
| | 3 | ✓ | ✓ | ✓ | | ✓ | |
| | 4 | | | ✓ | | | ✓ |
| | 5 | ✓ | ✓ | ✓ | ✓ | ✓ | |
| | 6 | ✓ | ✓ | ✓ | ✓ | | ✓ |
| | 7 | | | ✓ | | | |
| | 8 | | | ✓ | ✓ | | ✓ |
| | 9 | ✓ | ✓ | ✓ | | ✓ | ✓ |
| | 10 | | | ✓ | | | ✓ |
| | 11 | ✓ | ✓ | ✓ | | ✓ | |
| | 12 | | | ✓ | ✓ | | |
| | 13 | ✓ | ✓ | ✓ | ✓ | | ✓ |
| | 14 | ✓ | ✓ | ✓ | ✓ | ✓ | ✓ |
| | 15 | | ✓ | ✓ | ✓ | ✓ | ✓ |
| | 16 | ✓ | ✓ | | | ✓ | |
| | 17 | ✓ | ✓ | ✓ | ✓ | | ✓ |

Table (II) lists some of the studies identified in Subsection (IV-B). The table indicates which types of security issues or challenges are addressed by each study. In this paper, we attempt to focus only on research works that address security issues in the EC-assisted IoT paradigm. Table (III), on the other hand, shows the relationship between the security/privacy solutions and countermeasures discussed in Subsection (IV-A) on one side, and the security/privacy functions and services discussed in Subsection (IV-B) on the other side. This table illustrates which security countermeasure technique is addressing which security function and service type. We noticed that only a few numbers of studies tried to cover security issues associated with the EC-assisted IoT in particular. There are several issues for which current studies may not provide adequate solutions. Research in some security/privacy aspects of EC-assisted IoT still in progress, and many questions and problems are yet to be answered.

## V. Open Challenges and Future Research Directions

Although we have discussed the main security and privacy issues such as main mechanisms, attacks, and possible countermeasures, there still open emerging security/privacy challenges and issues that either not explained yet or need further exploration from an EC-assisted IoT paradigm perspective. This section extensively explains some of these open challenges and provides deep insights into some promising future research directions.

*1) Limited Device Capabilities:* existing IoT edge devices rely on compact battery-powered circuits with limited storage and computation capabilities. Therefore, they cannot support or implement conventional highly-secured, and sophisticated security techniques and schemes. This leads to the emergence of several weak links in the EC-assisted IoT networks, that can be exploited by intruders. Hence, a promising research direction could be to devise novel lightweight security/privacy schemes at different entities within the EC-assisted IoT infrastructure. For instance, designing lightweight middleware-based security management frameworks is one of these promising fields [85]. In addition, the existing trust management algorithms are complex and resource-consuming, and the tiny IoT edge devices can not support them. Thus, novel lightweight and compatible trust management algorithms must be devised for such IoT devices/nodes. Moreover, conventional cryptographic techniques and protocols need high computational powers, as they require a large encryption key size. Hence, they cannot be directly implemented in EC-assisted IoT network. This also shows the paramount importance to design new lightweight cryptographic techniques and protocols that possess small encryption keys and are deployable within the limited storage and CPU resources of EC-assisted IoT devices/nodes. Such lightweight cryptographic techniques should compromise between ensuring security and privacy on one side and satisfying the QoS requirements of time-sensitive EC-assisted IoT applications on the other side. Future research directions in this field include cryptographic schemes such as Elliptic Curve, Permutation-Based Lightweight, and Block-Ciphers Lightweight [86]. Furthermore, designing lightweight key exchange algorithms that ensure secure two-way communications in EC-assisted IoT networks is also a promising research direction.

*2) Comprehensive Trust Management Frameworks:* EC-assisted IoT networks are heterogeneous, as they are formed of different types of edge devices and various infrastructures. In addition, the ability of some edge nodes and servers to perform some complex processing tasks has encouraged developers to migrate trust modeling and evaluation from the cloud servers to edge nodes. Hence, multiple trust domains of multiple functional entities will coexist in EC-assisted IoT networks, which poses several open research challenges. Here we discuss some of them.

The heterogeneity of multiple trust domains at the network's edge must be carefully considered during the design of cryptographic schemes in order to enable efficient and distributed data encryption systems. Besides, authentication mechanisms need to specify a unique identity to each edge entity, as well as to support mutual authentication across all existing edge entities within the EC-assisted IoT network. Hence, and in order to address these issues, it is required to develop a dynamic and fine-grained multi-domain access control system that is aware of the cross-domain nature of the EC-assisted IoT network as well as inter-group hierarchical access control schemes.

It is also essential to develop efficient and dynamic privacy-preserving data update mechanisms from edge users' identity, interest, and location perspectives. In addition, it is required to develop trust establishment and evaluation frameworks for new edge entities in the EC-assisted IoT system that enable communication with new edge nodes/devices without the knowledge of third parties. Moreover, it is also imperative to develop dynamic and scalable trust evaluation mechanisms that consider several issues, such as updating trust values and tracking moving IoT edge devices. Furthermore, context-aware trust relationships based on social computing is also another issue that needs more investigation, exploration, and development from an EC-assisted IoT perspective.

It is also necessary to develop a universal and fine-grained trust management mechanism/model suitable for the heterogeneous EC-assisted IoT networks, as most of the conventional sophisticated trust management algorithms may not be able to be implemented directly within the limited-resource tiny IoT edge devices. Such a universal trust mechanism must support both scalability and mobility of the EC-assisted IoT ecosystem. Developing efficient and intelligent clustering mechanisms and algorithms based on trust management for the EC-assisted IoT paradigm is also a new research direction. Such mechanisms must be able to automatically detect and exclude malicious edge devices/nodes from the EC-assisted networks and hence ensuring system reliability and trust. Also, trust management mechanisms based on game theory is another new interesting research direction for the EC-assisted IoT paradigm [87].

*3) Mechanisms Orchestration and Standardization:* due to the massive software/hardware heterogeneity of the EC-assisted IoT ecosystem, it becomes imperative to efficiently orchestrate a various set of security and privacy schemes. This is done by developing flexible and unified security/privacy mechanisms, standards, platforms, and policies that support integrity, interoperability, heterogeneity, and show immunity against security threats. Developers and service providers must develop such unified security schemes taking into consideration the subtle operating specifications and differences of the underlying EC devices/nodes, as such details greatly impact the deployment and implementation of EC-assisted IoT infrastructure. Besides, taking into account that there are various third-party partners involved in developing EC-assisted IoT networks, such as network device vendors, application developers, and service providers, the problem of devising unified security and privacy schemes becomes even more challenging. Such parties should cooperate to develop interoperable security and privacy mechanisms in order to facilitate the flow of information with a high level of protection. Hence, security and privacy regulations are crucial in promoting the adaptation of secure EC-assisted IoT ecosystem.

*4) Authentication:* the explosive increase in the number and types of heterogeneous EC-assisted IoT nodes and devices make it crucial to ensure security and privacy across all edge nodes and interfaces. Towards this, efficient data integrity as well as flexible and scalable authentication and authorization mechanisms are necessary in order to meet the requirements of the growing and expanding EC-assisted IoT networks. One of the problems that needs more proper addressing is providing secure privacy-preserving authentication, auditing, and access control to system resources. Some edge users worry about keeping track of their actions or exposing their location or identity. Hence, solutions that provide secure access to the system and, at the same time, maintain the privacy of edge users are still open research problems that need more exploration and investigation from EC-assisted IoT perspective. For example, designing an identity-based mutual anonymous authentication key agreement protocols for the EC-assisted IoT paradigm would be a promising research direction. Also, utilizing hash chains and authenticated encryption [88] to develop lightweight authentication protocols that are able to provide security for EC-assisted IoT is another promising new research direction.

*5) Software Defined Networking (SDN) and Blockchain Techniques:* these technologies are grabbing considerable attention recently as they present innovative ideas for securing the distributed EC-assisted IoT architectures. In addition to its intelligent ability to reconfigure edge devices and route traffic of EC-assisted IoT networks, SDN also offers efficient and secure solutions for authentication and access control mechanisms [16], [37], [41], [89]. For example, developing lower computational delay and less communication resources SDN-based handover authentication management schemes for EC-assisted IoT is still one of the promising research directions [90], [91]. Also, distributed authentication based on SDN technology for EC-assisted IoT is another possible research direction.

Blockchain technology, on the other hand, can improve the security of the EC-assisted IoT paradigm as it permits only trusted IoT devices/nodes to interact with each other. Yet, there are still several promising open research directions. For example, developing security frameworks based on permissioned blockchain for the EC-assisted IoT paradigm is still an open research direction. In addition, due to the distributed nature of the edge nodes/devices in EC-assisted IoT ecosystem, decentralized security architectures based on hybrid SDN-blockchain is also one of the promising new research directions in the literature [91], [92]. In this architecture, the blockchain scheme is implemented to guarantee decentralized security to avoid a single point failure, whereas the SDN scheme is implemented to provide continuous monitoring of the EC-assisted IoT network. Utilizing blockchain to develop both authentication mechanisms and secure layer for edge devices/nodes in EC-assisted IoT, is also another promising research direction. Moreover, developing blockchain-based trusted data management schemes for cooperative authentication, authorization, and privacy-preserving in the EC-assisted IoT networks is also a new research direction [93], [94]. The integration of Ethereum blockchain architecture and artificial intelligence (AI) in order to enhance the security of EC-assisted IoT is also another interesting research direction [95]. Also, developing robust and lightweight optimization algorithms for the blockchain ecosystem is an open research challenge in access control and secure storage for the EC-assisted IoT paradigm.

*6) Data Issues:* security and privacy in data collection,

sharing, storage, and management are also still open research issues. Major research issues that need more exploration and investigation from an EC-assisted IoT perspective include mechanisms such as data confidentiality, integrity, privacy, etc. For example, the authors in [86] argue that Reliability, Availability, Integrity, and Nonrepudiation requirements all are not well addressed and investigated by any techniques in the literature from an EC-assisted IoT perspective. Therefore, they represent promising research directions. In addition, flexible, fine-grained, and self-adaptive data analytics schemes are also required in order to automatically identify the level of sensitivity of edge user data and provide the suitable security mechanisms to deal with it [39]. Furthermore, maintaining security and privacy to EC data storage is also one of the problems that needs to be addressed in the literature. Also, the problem of developing a mechanism to provide edge users with easy, safe, and secure access to distributed data storage and, at the same time, maintaining edge user privacy is still an open research direction.

Using traditional security methods, that are originally proposed for cloud servers, to protect data at the edge devices/nodes may not be feasible, given the huge difference between cloud servers and edge nodes in terms of computation and storage powers. Moreover, EC networks are distributed, scalable, and heterogeneous. This represents a challenge for security mechanisms that have to maintain efficiency and privacy for data storage, auditing, backup, and recovery.

Since edge devices are typically lightweight with limited computational capabilities and resources, it becomes imperative to device new lightweight schemes to perform secure data computation and processing. In particular, developing lightweight mechanisms to guarantee the correctness of data analytics while ensuring security is still a promising research direction in the EC-assisted IoT paradigm. This is due to the fact that edge users commonly migrate within the EC-assisted IoT network, and hence several edge servers might cooperatively serve a single edge user, which may result in mistakes in data analytics provided from/to edge servers. Thus, developing flexible and low-overhead provenance management techniques [96] for achieving a traceable and verifiable computation is also a promising research direction in the context of EC-assisted IoT.

On the other hand, since smart edge devices in EC-assisted IoT networks generate a massive amount of data at the network edge, it becomes imperative to incorporate both efficient data-sharing mechanisms and dynamic auto-update functions into the privacy-preserving schemes of EC-assisted IoT, which represents a possible future research direction. Furthermore, in order to reduce the quantity and availability of edge users' confidential data, it is essential to develop new techniques for distributing data processing amongst edge devices/nodes and transmitting only processed data at the different layers of the EC-assisted IoT system. Also, developing real-time systems for managing and orchestrating these distributed edge schemes and maintaining the correctness of data analytics becomes a crucial factor in deploying secure EC-assisted IoT infrastructure, which also needs more research and development.

*7) Joint Design:* it is also imperative to develop efficient security schemes that consider the joint design of mobility, handover, authentication, scalability, security, and/or privacy characteristics of EC-assisted IoT networks. As in such a paradigm, edge devices are frequently moving within the network's geographical area, or even rapidly joining and leaving the EC network. Hence, devising new real-time security mechanisms, such as authentication, access control, trust, etc., that can automatically and intelligently adapt to this rapid mobility and scalability of EC-assisted IoT network structure is also a very interesting future research direction.

*8) Machine Learning Techniques:* utilizing machine learning models, such as deep learning [97], reinforcement learning [98], [99], and deep reinforcement learning, to detect and predict malicious applications and adversarial activities at the EC level is also a new interesting research area for EC-assisted IoT systems [100], [101]. In particular, machine learning models can be exploited in developing intelligent security/privacy mechanisms and countermeasures. For example, they can be utilized in anomaly detection in order to ensure fine-grained authentication in EC-assisted IoT systems [37]. Also, they can be integrated with other techniques such as blockchain to provide, e.g., trust mechanisms for EC-assisted IoT, which represents a promising research direction.

On the other hand, since EC devices are becoming more heterogeneous in terms of available resources and software, this would make collaborative machine learning techniques more susceptible to exposing the training dataset of authorized participants. Hence, achieving secure and privacy-preserving data analysis in the EC-assisted paradigm based on distributed/federated learning strategies [102] without the leakage of the private training dataset is still an open research direction.

*9) Privacy and Extent of Hacked Data Usage:* future IoT devices are engaged in collecting and sharing information from various edge sensors ranging from environmental to user-related sensitive and private data (see Fig. (1)). As mentioned in [1] and the references therein, a plethora of unexpected privacy-sensitive information can be collected, such as daily routines, the number of residents, personal habits, etc. Attackers can collect this information by hacking homes' smart meters and edge devices. The question remains, what is the extent of private information that can be collected and extracted based on hacking non-critical data?

Preserving edge users' privacy by developing novel intelligent and lightweight data analytics mechanisms, which can automatically and adaptively identify the degree of sensitivity of edge user data, is a promising future research direction. For example, privacy-preserving for EC-assisted IoT based on techniques such as Privacy by Design (PbD), Software Defined Privacy (SDP), and SDN-based privacy-preserving routing is a possible research direction. Although some of these privacy-preserving concepts have been proposed for the traditional IoT paradigm, they can be further extended and enhanced to support the EC-assisted IoT paradigm, taking into account the new features of this paradigm that we have discussed in this paper.

## VI. Conclusion

This paper presents a comprehensive survey on security and privacy issues for the EC-assisted IoT paradigm. To achieve this goal, we first provide an overview of EC-assisted IoT including its applications and architecture. Then, we discuss the advantages and limitations of integrating EC and IoT paradigms. After that, we conduct an in-depth analysis of security and privacy in the context of EC-assisted IoT. In particular, we extensively survey the key classifications and types of possible IoT network security and privacy attacks and the corresponding countermeasures at different IoT network layers along with the related research works. After that, we provide analysis of security and privacy mechanisms, then we classify some of the security and privacy issues reported in the existing research works based on security services and based on security objectives and functions. Lastly, open security-related research issues and challenges, in the context of EC-assisted IoT, are extensively provided along with possible research directions.